\documentclass[12pt]{article}

\usepackage{slashed}
\begin{document}

\title{Families from Supergroups and Predictions for Leptonic CP Violation}
\author{{\bf S.M. Barr and Heng-Yu Chen} \\
Department of Physics and Astronomy \\ 
and \\
Bartol Research Institute \\ University of Delaware
Newark, Delaware 19716} \maketitle

\begin{abstract}
As was shown in 1984 by Caneschi, Farrar, and Schwimmer, decomposing representations of the supergroup $SU(M|N)$, can give interesting anomaly-free sets of fermion representations of $SU(M) \times SU(N) \times U(1)$. It is shown here that such groups can be used to construct realistic grand unified models with non-abelian gauged family symmetries. A particularly simple three-family example based on $SU(5) \times SU(2) \times U(1)$ is studied. The forms of the mass matrices, including that of the right-handed neutrinos, are determined in terms of $SU(2)$ Clebsch coefficients; and the model is able to fit the lepton sector and predict the Dirac CP-violating phase of the neutrinos. Models of this type would have a rich phenomenology if part of the family symmetry is broken near the electroweak scale. 
\end{abstract}

\section{Introduction} 

One way of finding chiral sets of fermions that are anomaly-free under product gauge groups is to decompose anomaly-free multiplets of larger  groups. For example, by decomposing the ${\bf 10} + \overline{{\bf 5}}$ of $SU(5)$ under its $SU(3) \times SU(2) \times U(1)$ subgroup, one finds the anomaly-free set that comprises the fermions of the Standard Model. And by decomposing the ${\bf 16}$ of $SO(10)$ under its $SU(5) \times U(1)$ subgroup, one finds the anomaly-free set
${\bf 10}^1 + \overline{{\bf 5}}^{-3} + {\bf 1}^5$. 

It was shown in \cite{CFS} that interesting sets of fermions that are anomaly-free under groups of the form $SU(M) \times SU(N) \times U(1)$ can be found by decomposing multiplets of the supergroup $SU(M|N)$ \cite{supergroups}. The idea is based on the fact that the Casimirs of $SU(M|N)$ only depend on $(M-N)$. Thus the third-order Casimirs for the groups $SU(M+P|P)$ are the same for any $P$, and thus the same as for $SU(M)$. If one considers, therefore, an irreducible fermion representation that is anomaly-free (i.e. has vanishing third-order Casimir) under $SU(M)$, the corresponding Young tableaux representation of $SU(M+P|P)$ will yield anomaly-free sets when decomposed under the bosonic subgroup $SU(M+P) \times SU(P) \times U(1)$. 

To take a simple example, consider the totally antisymmetric rank-four tensor multiplet of $SU(4)$. This is a singlet, and thus trivially anomaly free. Consequently, the representation with the same Young tableau yields anomaly-free sets of fermions when decomposed under the $SU(4 +P) \times SU(P) \times U(1)$ subgroup of $SU(4 + P|P)$. This decomposition gives the multiplets

\begin{equation}
([4],(0))^{P} + (\overline{[3]}, \overline{(1)})^{-(P+1)} 
+ ([2],(2))^{(P+2)} + (\overline{[1]}, \overline{(3)})^{-(P+3)}
+ ([0],(4))^{P+4},
\end{equation}

\noindent where $[m]$ and $(m)$ stand for the rank-$m$ tensor multiplets that are totally antisymmetric and totally symmetric in the indices, respectively, and the overbar stands for the conjugate multiplets. The superscripts are the $U(1)$ charges. 
If we take $P=1$, this gives the anomaly-free set of $SU(5) \times U(1)$
multiplets $\overline{{\bf 5}}^1 + {\bf 10}^{-2} + {\bf 10}^3 + \overline{{\bf 5}}^{-4} + {\bf 1}^5$. If we take $P=2$, this gives the anomaly-free set of 
$SU(6) \times SU(2) \times U(1)$
multiplets $(\overline{{\bf 15}}, {\bf 1})^2 + ({\bf 20}, {\bf 2})^{-3} + ({\bf 15}, {\bf 3})^4 + (\overline{{\bf 6}}, {\bf 4})^{-5} + ({\bf 1}, {\bf 5})^5$.

The anomaly-free sets constructed in \cite{CFS} are interesting from the point of view of gauged family symmetry. In a theory having the gauge group $SU(M) \times SU(N) \times U(1)$, the first factor could contain the Standard Model group if $M\geq 5$, while $SU(N)$ could be a family group if it has three-dimensional representations, whether irreducible or reducible.   

An anomaly-free set of fermions that contains exactly three families of the Standard Model can be obtained by looking at the rank-3 tensors of $SU(3 +P|P)$. This gives the anomaly free-set of $SU(3 + P) \times SU(P) \times U(1)$ fermion multiplets

\begin{equation}
(\overline{[3]}, \overline{(0)})^{-P} 
+ ([2],(1))^{(P+1)} + (\overline{[1]}, \overline{(2)})^{-(P+2)}
+ ([0],(3))^{P+3},
\end{equation} 
 
\noindent An interesting case, which gives a family group $SU(3)$, is obtained by setting $P=3$ in Eq. (2), in which case the group is $SU(6) \times SU(3) \times U(1)$ 
and the multiplets are $({\bf 20}, {\bf 1})^{-3} + 
({\bf 15}^{-3}, {\bf 3})^4 + (\overline{{\bf 6}}, {\bf 6})^{-5} 
+({\bf 1}, {\bf 10})^6$. This contains in addition to the Standard Model fermions many fermions that are vector-like under the Standard Model group. Under $SU(5)$, it contains in addition to the three families of ${\bf 10} + \overline{{\bf 5}}$, three sets of ${\bf 5} + \overline{{\bf 5}}$, one of 
${\bf 10} + \overline{{\bf 10}}$, and sixteen singlets. 

A more economical case is obtained by setting $P =2$ in Eq. (2). Then one has the following group and fermion multiplets:

\begin{equation}
SU(5) \times SU(2) \times U(1): \;\;\;\;\;\;  ({\bf 10}, {\bf 1})^{-2} + 
({\bf 10}, {\bf 2})^3 + (\overline{{\bf 5}}, {\bf 3})^{-4} + ({\bf 1}, {\bf 4})^5.
\end{equation}

\noindent The only fermions this contains besides those of the Standard Model (SM) are four SM-singlets, which can play the role of the right-handed neutrinos. This is the simplest and most economical case based on the constructions of \cite{CFS}. We shall therefore study it in detail. As will be seen, the family $SU(2)$ gives non-trivial forms for the fermion mass matrices, fits the lepton sector well, and predicts the Dirac CP phase of the neutrinos. 

\section{The Minimal $SU(5) \times SU(2) \times U(1)$ Model}

With the fermions multiplets given in Eq. (3), all fermion masses can be obtained with just four Higgs multiplets (which will be distinguished from fermion multiplets by a subscript $H$):

\begin{equation}
({\bf 5}, {\bf 1})^4_H, \;\; ({\bf 5}, {\bf 2})^{-1}_H, \;\; 
({\bf 5}, {\bf 3})^{-6}_H, \;\; ({\bf 1}, {\bf 3})^{-10}_H.
\end{equation}

\noindent These Higgs multiplets have the following Yukawa couplings to the fermions:

\begin{equation}
\begin{array}{lcl}
u \; {\rm masses \; from} \;\; {\bf 10}\; {\bf 10} \; {\bf 5}_H \; {\rm terms:} \;\;\; & & a\; ({\bf 10}, {\bf 1})^{-2} ({\bf 10}, {\bf 1})^{-2}
\;\; ({\bf 5}, {\bf 1})^4_H \\ 
& + & b \; ({\bf 10}, {\bf 1})^{-2} ({\bf 10}, {\bf 2})^{3}
\;\; ({\bf 5}, {\bf 2})^{-1}_H \\ 
& + & c \; ({\bf 10}, {\bf 2})^{3} ({\bf 10}, {\bf 2})^{3}
\;\; ({\bf 5}, {\bf 3})^{-6}_H, \\ & & \\
d, \ell^- \; {\rm masses \; from} \;\; {\bf 10}\; \overline{{\bf 5}} \; \overline{{\bf 5}}_H \; {\rm terms:}& & e \; ({\bf 10}, {\bf 1})^{-2} (\overline{{\bf 5}}, {\bf 3})^{-4}
\;\; [({\bf 5}, {\bf 3})^{-6}_H]^* \\ 
& + &  f \; ({\bf 10}, {\bf 2})^3 (\overline{{\bf 5}}, {\bf 3})^{-4}
\;\; [({\bf 5}, {\bf 2})^{-1}_H]^* \\ & & \\
\nu \; {\rm Dirac \; masses \; from} \;\; \overline{{\bf 5}}\; {\bf 1} \; {\bf 5}_H \; {\rm terms:}& & 
g \; (\overline{{\bf 5}}, {\bf 3})^{-4} ({\bf 1}, {\bf 4})^5
\;\; ({\bf 5}, {\bf 2})^{-1}_H \\ & & \\
\nu^c \; {\rm masses\; from} \;\; {\bf 1}\; {\bf 1} \; {\bf 1}_H \; {\rm terms:} & &  
h \; ({\bf 1}, {\bf 4})^5 ({\bf 1}, {\bf 4})^5
\;\; ({\bf 1}, {\bf 3})^{-10}_H.
\end{array}
\end{equation}

\noindent Note that an $SU(2)$-singlet mass for the right-handed neutrinos, 
{\it i.e.} $({\bf 1}, {\bf 4})^5 {\bf 1}, {\bf 4})^5 ({\bf 1}, {\bf 1})_H^{-10}$, is forbidden by Fermi statistics, since the symmetric product
of two 4-plets of $SU(2)$ does not contain a singlet. 

As we shall see in detail, the forms of mass matrices of the quarks and leptons that arise from the Yukawa terms in Eq. (5) are determined by the $SU(2)$ family symmetry, and the Clebsch coefficients of $SU(2)$. 
 
Let us denote the vacuum expectation values (VEVs) of the Higgs multiplets shown in Eq. (4) as follows

\begin{equation}
\begin{array}{ccl}
\langle ({\bf 5}, {\bf 1})^4_H \rangle & = & S, \\ & & \\
\langle ({\bf 5}, {\bf 2})^{-1}_H \rangle & = & (d_{\downarrow}, d_{\uparrow}), \\ & & \\ 
\langle ({\bf 5}, {\bf 3})^{-6}_H \rangle & = & (v_1, v_2, v_3), \\ & & \\
\langle ({\bf 1}, {\bf 3})^{-10} \rangle &  =  &  (t_1, t_2, t_3).
\end{array}
\end{equation}

\noindent Here we have expressed the VEVs of the $SU(2)$ triplets in a ``Cartesian 
basis". But we can also denote them in a ``spherical basis", with  
$v_{\pm} \equiv (v_1 \pm i v_2)/\sqrt{2}$, $v_0 \equiv v_3$, and 
$t_{\pm} \equiv (t_1 \pm i t_2)/\sqrt{2}$,
$t_0 \equiv t_3$. The VEV $(t_1,t_2,t_3)$ is a complex vector. If we assume that its real and imaginary parts are aligned, then we can choose the basis in $SU(2)$ space so that $(t_1, t_2, t_3) = (0,0,t)$. Such alignment happens if a certain quartic self-coupling of $({\bf 1}, {\bf 3})^{-10}_H$ has the right sign. The most general renormalizable potential for this field is of the form $V(\vec{t}) = \mu^2 (\vec{t}^* \cdot \vec{t}) + 
\lambda (\vec{t}^* \cdot \vec{t})^2 + \lambda' (\vec{t} \cdot \vec{t})
(\vec{t}^* \cdot \vec{t}^*) + \lambda'' ( \vec{t}^* \times \vec{t})^2$. 
If we write $\vec{t} = \vec{a} + i \vec{b}$, where $\vec{a}$ and $\vec{b}$ are real vectors, then $V = \mu^2 (a^2 + b^2) + (\lambda + \lambda') (a^2 + b^2)^2 
- 4 (\lambda' + \lambda'') a^2 b^2 \sin^2 \theta_{ab}$. There are two cases: Case I with $\lambda' + \lambda'' < 0$, and Case II with $\lambda' + \lambda'' >0$. In Case I, the angle $\theta_{ab}$ between $\vec{a}$ and $\vec{b}$ vanishes. Then $\vec{t} = \hat{a} (a+ib)$, where the phase of $a+ib$ can be gauged away. Choosing $\hat{a}$ to point in the 3 direction, and defining $t \equiv \sqrt{a^2 + b^2}$, one ends up with the form $(t_1, t_2, t_3) = (0,0,t)$. In Case II, $\theta_{ab} = \pi/2$, so 
$\vec{a}$ and $\vec{b}$ are perpendicular to each other, and one can choose the basis in $SU(2)$ space 
so that $(t_1, t_2, t_3) = (0, it', t)$. Moreover, in this case the term with  $\sin^2  \theta_{ab}$  becomes $-|\lambda' + \lambda''| a^2 b^2$, meaning that $a$ and $b$ become of equal magnitude, and
one has $(t_1, t_2, t_3) = (0, it, t)$. These two cases give different mass matrices for the right-handed neutrinos and will both be examined below.

We will define the complex numbers

\begin{equation}
x \equiv d_{\downarrow}/d_{\uparrow}, \;\;\;\;  z_1 \equiv v_1/v_3, 
\;\;\;\; z_2 \equiv v_2/v_3.
\end{equation}

Let us similarly denote the fermion multiplet $(\overline{{\bf 5}}, {\bf 3})^{-4}$ by 
$(\overline{{\bf 5}}_1, \overline{{\bf 5}}_2, \overline{{\bf 5}}_3)$ or
$(\overline{{\bf 5}}_-, \overline{{\bf 5}}_0, \overline{{\bf 5}}_+)$ and the 
fermion multiplet $({\bf 10}, {\bf 2})^3$ by $({\bf 10}_{\downarrow}, {\bf 10}_{\uparrow})$.
The $({\bf 10}, {\bf 1})^{-2}$ we will denote simply by ${\bf 10}$, without any subscript. Then the $3 \times 3$ mass matrix of the up quarks mass can be written

\begin{equation}
\left( {\bf 10}_{\downarrow}, {\bf 10}_{\uparrow}, {\bf 10} \right)_{u}
\left( \begin{array}{ccc} c v_+ & cv_0/\sqrt{2} & b d_{\uparrow}/2 \\
cv_0/\sqrt{2} & cv_- & - b d_{\downarrow}/2 \\
bd_{\uparrow}/2 & - b d_{\downarrow}/2 & a S \end{array} 
\right) \left( \begin{array}{c} {\bf 10}_{\downarrow} \\
{\bf 10}_{\uparrow} \\ {\bf 10} \end{array} \right)_{u^c},
\end{equation}
 
\noindent so that the up quark mass matrix can be written in the form

\begin{equation}
M_u = \mu_u \; \left( \begin{array}{ccc}
\delta (z_1 + i z_2) & \delta & \epsilon \\
\delta & \delta (z_1 - i z_2) & - \epsilon x \\
\epsilon & -\epsilon x & 1 \end{array} \right),
\end{equation}

\noindent where $\mu_u = aS$, $\epsilon = \frac{b d_{\uparrow}}{2aS}$,  
$\delta = \frac{cv_0}{\sqrt{2} aS}$. 

Since the VEVs that give the fermions mass do not break $SU(4)_c$, one obtains the unrealistic ``minimal $SU(5)$" relation \cite{GG} between the down quark and charged lepton mass matrices: $M_d = M_{\ell}^T$. These come from the term

\begin{equation}
\left( {\bf 10}_{\downarrow}, {\bf 10}_{\uparrow}, {\bf 10} \right)_{d \; ({\rm or}\; \ell^c)}
\left( \begin{array}{ccc} fd_{\uparrow}^*/\sqrt{3} & i f d_{\uparrow}^*/\sqrt{3} & fd_{\downarrow}^*/\sqrt{3} \\
- f d_{\downarrow}^*/\sqrt{3} & i f d_{\downarrow}^*/\sqrt{3}  & f d_{\uparrow}^*/\sqrt{3} \\
e v_1^* & e v_2^* & e v_3^* \end{array} 
\right) \left( \begin{array}{c} \overline{{\bf 5}}_1 \\
\overline{{\bf 5}}_2 \\ \overline{{\bf 5}}_3 \end{array} \right)_{d^c \; ({\rm or} \; \ell)},
\end{equation}
 
\noindent This gives 

\begin{equation}
M_d = \mu_d \left( \begin{array}{ccc} \eta & i\eta & \eta x^*  \\ -\eta x^*  & i \eta x^*  & \eta \\
z_1^* & z_2^* & 1 \end{array} \right), \;\;\;\; 
M_{\ell} = \mu_d \left( \begin{array}{ccc} \eta & -\eta x^* & z_1^* \\ i \eta & i \eta x^*  & z_2^* \\
\eta x^*   & \eta & 1 \end{array} \right),
\end{equation}
 
\noindent where $\mu_d = e v_0^*$. 
$\eta = \frac{f d_{\uparrow}^*}{\sqrt{3} e v_0^*}$, and we have used the complex parameters $x$, $z_1$, and $z_2$ parameters in Eq. (7).

The neutrino mass matrix arises through a Type I see-saw mechanism \cite{TypeI}. There are three left-handed neutrinos in the $(\overline{{\bf 5}}, {\bf 3})^{-4}$ and four left-handed anti-neutrinos in the $({\bf 1}, {\bf 4})^5$. The Dirac neutrino mass matrix comes from

\begin{equation}
\left( \overline{{\bf 5}}_+, \overline{{\bf 5}}_0, \overline{{\bf 5}}_- \right)_{\nu} 
\left( \begin{array}{cccc} 0 & 0 & -\sqrt{\frac{1}{3}} g d_{\downarrow} & g d_{\uparrow} \\
0 &  \sqrt{\frac{2}{3}} g d_{\downarrow} & - \sqrt{\frac{2}{3}} d_{\uparrow} & 0
\\ -g d_{\downarrow} & \sqrt{\frac{1}{3}} g d_{\uparrow} & 0 & 0 \end{array} 
\right) \left( \begin{array}{c} {\bf 1}_{3/2} \\
{\bf 1}_{1/2} \\ {\bf 1}_{-1/2} \\ {\bf 1}_{-3/2} \end{array} \right)_{\nu^c},
\end{equation}

\noindent where the form is entirely determined by $SU(2)$ Clebsch coefficients.
The $4\times 4$ Majorana mass matrix of the $\nu^c$ is also determined by Clebsch coefficients. 
In Case I, where $(t_-, t_0, t_+) = (0,t,0)$, one then finds 

\begin{equation}
\left( {\bf 1}_{3/2}, {\bf 1}_{1/2}, {\bf 1}_{-1/2}, {\bf 1}_{-3/2} \right)_{\nu^c} 
\left( \begin{array}{cccc} 0 & 0 & 0 & \frac{3}{\sqrt{20}} h t \\
0 & 0 & - \frac{1}{\sqrt{20}} h t & 0
\\ 0 & - \frac{1}{\sqrt{20}} h t & 0 & 0 \\ \frac{3}{\sqrt{20}} h t
& 0 & 0 & 0 \end{array} 
\right) \left( \begin{array}{c} {\bf 1}_{3/2} \\
{\bf 1}_{1/2} \\ {\bf 1}_{-1/2} \\ {\bf 1}_{-3/2} \end{array} \right)_{\nu^c},
\end{equation}

\noindent From the see-saw formula $M_{\nu} = - M_{Dirac} M_R^{-1} M_{Dirac}^T$, one finds

\begin{equation}
 \left[ - 2 \sqrt{5} \frac{g^2 d_{\uparrow}^2}{3ht} \right]
\left( \nu_+, \nu_0, \nu_- \right) \left(
\begin{array}{ccc} 0 & \sqrt{2} x^2 & 0 \\
\sqrt{2} x^2 & 4 x & \sqrt{2} \\
0 & \sqrt{2} & 0 \end{array} \right) \left( 
\begin{array}{c} \nu_+ \\ \nu_0 \\ \nu_- \end{array} \right).
\end{equation}  

\noindent Writing this in the Cartesian basis $(\nu_1, \nu_2, \nu_3)$, and defining
$\mu_{\nu} = -\frac{2 \sqrt{5} g^2 d_{\uparrow}^2}{3 ht}$, one has

\begin{equation}
\mu_{\nu}
\left( \nu_1, \nu_2, \nu_3 \right) \left(
\begin{array}{ccc} 0 & 0 & x^2 +1  \\
0 & 0  & i (x^2 - 1) \\
x^2 +1 & i (x^2 - 1) & 1 \end{array} \right) \left( 
\begin{array}{c} \nu_1 \\ \nu_2 \\ \nu_3 \end{array} \right).
\end{equation}  

\noindent This is not, however, the most general form of the neutrino mass matrix,
because another operator can contribute to it, namely the effective dim-5 operator
$(\overline{{\bf 5}}, {\bf 3})^{-4} (\overline{{\bf 5}}, {\bf 3})^{-4} 
({\bf 5}, {\bf 1})^4_H ({\bf 5}, {\bf 1})^4_H$. In the Cartesian basis, this just gives the identity matrix. Defining the ratio of the coefficient of this term to 
$\mu_{\nu}$ by the complex number $y$, we have

\begin{equation}
M_{\nu} = \mu_{\nu}
\left( \begin{array}{ccc} y & 0 & x^2 +1  \\ & & \\
0 & y  & i (x^2 - 1) \\ & & \\
x^2 +1 & i (x^2 - 1) & 4x+y \end{array} \right).
\end{equation}  

\noindent Note that the complex parameter $y$ actually makes a difference for the neutrino mixing angles and mass splittings, despite appearing as the coefficient of the identity matrix. This is so, because $M_{\nu}$ is complex and {\it symmetric} and thus diagonalized by $U_{\nu} M_{\nu} U^T_{\nu}$ rather than by  
$U_{\nu} M_{\nu} U^T_{\nu}$. 

For Case II, where $(t_1, t_2, t_3) = (0,it,t)$, one has $(t_-, t_0, t_+) =
(-t/\sqrt{2}, t, +t/\sqrt{2})$, This gives the following mass matrix for the right-handed neutrinos:

\begin{equation}
\left( {\bf 1}_{3/2}, {\bf 1}_{1/2}, {\bf 1}_{-1/2}, {\bf 1}_{-3/2} \right)_{\nu^c} 
\left( \begin{array}{cccc} 0 & 0 & -\frac{\sqrt{3}}{\sqrt{20}} ht & \frac{3}{\sqrt{20}} h t \\
0 & \frac{1}{\sqrt{5}} ht & - \frac{1}{\sqrt{20}} h t &  \frac{\sqrt{3}}{\sqrt{20}} ht
\\ -\frac{\sqrt{3}}{\sqrt{20}} ht & - \frac{1}{\sqrt{20}} h t & -\frac{1}{\sqrt{5}} ht & 0 \\ \frac{3}{\sqrt{20}} h t
& \frac{\sqrt{3}}{\sqrt{20}} ht & 0 & 0 \end{array} 
\right) \left( \begin{array}{c} {\bf 1}_{3/2} \\
{\bf 1}_{1/2} \\ {\bf 1}_{-1/2} \\ {\bf 1}_{-3/2} \end{array} \right)_{\nu^c},
\end{equation}

\noindent After straightforward algebra, this gives the following mass matrix for the three light neutrinos in a Cartesian basis:

\begin{equation}
M_{\nu} = \frac{1}{2} \mu_{\nu}
\left( \begin{array}{ccc} y & -i (x^2 +1) & x^2 + 1  \\ & & \\
-i(x^2 +1 ) & y - 4x  & i (1 -2x -x^2)  \\ & & \\
x^2 +1 & i (1 -2x - x^2) & y + 2(x^2 -1) \end{array} \right),
\end{equation}  

\noindent which is to be compared to Eq. (16).

\noindent The model described above is very simple, but is not fully realistic, because it gives the ``minimal $SU(5)$ relation" $M_d|_{M_{GUT}} = M_{\ell}|_{M_{GUT}}$ \cite{GG}.  Thus, if it fits the masses of the charged leptons, it will get the masses of the down quarks wrong by factors of order one. This defect could be repaired by introducing more Higgs fields into the model. It is interesting, however, to examine how close this minimal model comes to being realistic in other respects. Let us therefore see how well it fits the other quark and lepton parameters besides $m_d$, $m_s$ and $m_b$. 

The three parameters $\mu_u$, $\mu_d$ and $\mu_{\nu}$ that appear in Eqs. (9), (11), (16) and (18) allow one to fit the overall scales of the up quark, charged lepton, and neutrino masses. That leaves the inter-family mass ratios and the mixing parameters to be accounted for.

The most striking feature of the observed inter-family fermion mass ratios is that they are hierarchical. That can partly be explained in this model by the fact that the three families are distinguished from each other by how they transform under the $SU(2)$ family symmetry.  For instance, because of $SU(2)$, three different types of Higgs multiplet contribute to the up quark masses, as one sees from Eq. (5). If one assumes a hierarchy among the VEVs (or Yukawa coefficients, or both) of those three Higgs multiplets, one can have $\delta \ll \epsilon \ll 1$, which gives $m_u \ll m_c \ll m_t$, as is apparent from Eq. (9).  Two types of Higgs multiplets contribute to the down quark (and charged lepton) masses, as shown in Eq. (5). If one assumes a hierarchy in their VEVs (or Yukawa couplings, or both), one can have $\eta \ll 1$.  This would explain why the third family of down quarks and charged leptons is heavier than the first two families, as Eq. (11) shows.   However, it would not explain the lightness of the first family compared to the second for the down quarks and charged leptons.  As one can see from Eq. (11), that would require a certain relationship (which will be given later) to hold among the parameters $x$, $z_1$ and $z_2$. 

Each of the parameters $x$, $z_1$ and $z_2$ is defined as a ratio of VEVs of different components of an $SU(2)$ Higgs multiplet. One would therefore naturally expect that these (complex) parameters would have magnitudes of $O(1)$.  Thus, the relationship among them that would make $m_e/m_{\mu} \ll 1$ and $m_d/m_s \ll 1$ would involve a fine-tuning of order $10^{-2}$. 
 
The hierarchies $\delta \ll \epsilon \ll 1$ and $\eta \ll 1$ would also partially explain the smallness of the CKM angles.  An examination of Eqs. (9) and (11) shows that $V_{cb}$ and $V_{ub}$ come out to be of order $\eta$, while the Cabibbo mixing $V_{us}$ comes out to be $O(1)$ if $x$, $z_1$ and $z_2$ are arbitrary parameters of $O(1)$.  The ``fine-tuning" required to fit the Cabibbo angle is mild, but a tuning of order $10^{-1}$ is required to explain the smallness of 
$|V_{ub}|$. This tuning takes the form of a relation among $x$, $z_1$ and $z_2$ that must be approximately satisfied. 

If the parameters $x$, $z_1$ and $z_2$ have magnitudes of $O(1)$, as one would naturally expect, then the forms of the lepton mass matrices given in Eq. (11) and Eqs. (16) and (18) show that the PMNS angles should typically be of $O(1)$ as well, and that the ratio of neutrino masses should not be small. Thus, this model can account in a natural way for most of the qualitative features of the quark and lepton mass ratios and mixing angles. The two exceptions are the smallness of $m_e/m_{\mu}$ (and $m_d/m_s$) and the smallness of $|V_{ub}|$, each of which requires a somewhat tuned condition to hold among the complex parameters $x$, $z_1$, and $z_2$.

\section{The Lepton Sector in the Minimal Model}

Let us now see whether the simple model we have presented can fit the lepton sector, {\it i.e.} the masses of the charged leptons and neutrinos, and the PMNS angles.

As noted before, the fact that $m_e \ll m_{\mu}$ requires a tuning of parameters. As can be seen from an inspection of Eq. (11), for $|z_1|$, 
$|z_2|$ and $|x|$ of $O(1)$, and $|\eta|$ small, the three eigenvalues of $M_{\ell}$ are of order $|\mu_d|$, $|\eta \mu_d|$, and $|\eta \mu_d|$. 
To have $m_e \sim 10^{-2} m_{\mu}$ requires that $|\det M_{\ell}| \sim
10^{-2} |\eta^2 \mu_d^3|$. This yields the condition that 

\begin{equation}
\left| 1 - \frac{x^2 - 1}{2x} z_1 + i \frac{x^2 + 1}{2x} z_2 \right| \sim 10^{-2}.
\end{equation}

\noindent It will make no significant difference, and will simply calculations, if in fitting the neutrino properties we simply set this small quantity to zero. In that case, solving a quadratic equation allows one to solve for $x$ in terms of $z_1$ and $z_2$:

\begin{equation}
x \cong \frac{1 \pm \sqrt{1 + z_1^{\;2} + z_2^{\; 2}}}{z_1 + i z_2}.
\end{equation}

Suppose the mass matrices $M_{\ell}$ and $M_{\nu}$ are diagonalized by the following unitary transformations: $U_{\ell} M_{\ell} V_{\ell}^{\dag} =
M_{\ell}^{diagonal}$ and $U_{\nu} M_{\nu} U_{\nu}^T =
M_{\nu}^{diagonal}$. Then, with our conventions, the PMNS matrix is given by 
$U_{PMNS} = U_{\ell}^* U_{\nu}^T$. If we ignore effects that are subleading by order 
$|\eta|^2$, the unitary matrix $U_{\ell}$ depends only on the complex parameters $z_1$, $z_2$, and $x$, as can be seen by inspection of the form of $M_{\ell}$ given in Eq. (11). (The matrix $V_{\ell}$ depends on $\eta$ in leading order, but does not contribute to $U_{PMNS}$.) In fact it is easy to write an explicit form of $U_{\ell}$:

\begin{equation}
U_{\ell} = \left( \begin{array}{ccc} \cos \theta_{\ell 12} & - \sin \theta_{\ell 12} & 0 \\ (\sin \theta_{\ell 12})^* & (\cos \theta_{\ell 12})^* & 0 \\ 0 & 0 & 1 \end{array} \right) \left( \begin{array}{ccc}
\frac{z_2^*}{N_{12}} & -\frac{z_1^*}{N_{12}} & 0 \\ \frac{z_1}{N N_{12}}
& \frac{z_2}{N N_{12}} & - \frac{N_{12}}{N} \\ \frac{z_1}{N} &
\frac{z_2}{N} & \frac{1}{N} \end{array} \right),
\end{equation}

\noindent where $N_{12} \equiv \sqrt{|z_1|^2 + |z_2|^2}$, 
$N \equiv \sqrt{1 + N_{12}^2} = \sqrt{1 + |z_1|^2 + |z_2|^2}$, and 

\begin{equation}
\frac{\sin \theta_{\ell 12}}{\cos \theta_{\ell 12}}
\equiv \frac{i (z_1^{*2} + z_2^{*2}) \sqrt{1 + |z_1|^2 + |z_2|^2}}{ 
|z_1 - i z_2|^2 + (|z_1|^2 + |z_2|^2) ( -1 \pm \sqrt{1 + z_1^{*2}
+ z_2^{*2}})},
\end{equation}

\noindent where we have used Eq. (20) to eliminate the parameter $x$ and 
write $U_{\ell}$ entirely in terms of $z_1$ and $z_2$. 

The diagonalization of $M_{\nu}$, given in Eq. (16) for Case I and Eq. (18) for Case II, must be done numerically. This requires searching over three complex parameters of $O(1)$, namely $z_1$, $z_2$, and $y$. For each choice of these parameters, one can compute the PMNS angles and the ratio of neutrino mass splittings $\Delta m^2_{12}/\Delta m^2_{23}$. (The overall scale of the neutrino masses is set by the parameter $\mu_{\nu}$.)
One might think that one should be able to fit these four experimental numbers with the three complex model parameters $z_1$, $z_2$, and $y$. A good fit is not guaranteed to exist, however, as the equations are nonlinear. 

For Case I, we have done a numerical search of parameter space and found that there are values of the parameters that give excellent fits to the three PMNS angles, but none of them also gives a small enough value for the ratio of mass splittings $\Delta m^2_{12}/\Delta m^2_{23}$. 

For Case II, we have two found satisfactory solutions for the leptons, one corresponding the minus
sign in Eq. (20), and the other corresponding to the plus sign. We will call these Solutions 1 and 2, respectively. These two solutions give a good fit fit all three neutrino mixing angles and the ratio of neutrino mass splittings $\Delta m^2_{12}/\Delta m^2_{23}$, but give different predictions for the Dirac CP phase of the neutrinos $\delta_{CP}$.

In Table I, we present the fits to the neutrino mixing angles and the predictions of $\delta_{CP}$ for the two solutions. These were found in the following way. We searched over the three complex parameters $z_1$, $z_2$, $y$ and kept only those points which yielded values for the three PMNS angles and for the ratio of neutrino mass splittings that were each within one-sigma of the experimental value. The error bars in the second and third columns of Table I represent the standard deviation of the values obtained in this way. One notes that the prediction for the Dirac CP phase of the neutrinos $\delta_{CP}$ is fairly sharp for each of the two solutions.
The fourth column in Table I gives the $1\sigma$ best fit values from the 2014 particle data group \cite{pdg2014}, and the fifth column gives the best fit values from the 2016 particle data group \cite{pdg2016}. In Table II, we give the values of the complex model parameters $z_1$, $z_2$ and $y$ for the two solutions.

\vspace{2cm}

\noindent {\bf Table I.} The values of the PMNS parameters for the two solutions of Case II.

\[
\begin{tabular}{|l|l|l|l|l|}
\hline Quantity & Solution 1 &Solution 2 & $1 \sigma$ best fit$^5$ & best fit$^6$ \\
\hline \hline $\sin^2 \theta_{12}$ & $0.321 \pm 0.004 $ & 
$0.307 \pm 0.011$ &  $0.308 \pm 0.017$ & $0.297$ \\  
\hline $\sin^2 \theta_{23}$ & $0.467 \pm 0.0026$ & $0.457 \pm 0.0065$ & $0.437^{+0.033}_{-0.023}$
& $0.437$ \\
\hline $\sin^2 \theta_{13}$ & $0.0231 \pm 0.001$ & $0.0234 \pm 0.0015$ & $0.0234^{+0.002}_{-0.0019}$
& $0.0214$ \\
\hline $\delta_{CP}$ (rad) & $0.829 \pm 0.0035$ & $-0.617 \pm 0.0047$ & & \\
\hline $\delta_{CP}/\pi$ & $0.264 \pm 0.0011$ & $-0.196 \pm 0.0015$ & & \\
\hline 
\end{tabular}
\]

\noindent {\bf Table II.} The values of the complex parameters $z_1$, $z_2$, and $y$ for the two solutions for Case II. 

\[
\begin{tabular}{|l|l|l|}
\hline Quantity & Solution 1 & Solution 2 \\
\hline \hline Re$(z_1)$ & $1.51 \pm 0.004$ & $-0.098 \pm 0.006$ \\
\hline Im$(z_1)$ & $-0.064 \pm 0.009$ & $-1.19 \pm 0.0035$ \\
\hline Re$(z_2)$ & $0.13 \pm 0.008$ & $-0.056 \pm 0.0028$ \\
\hline Im$(z_2)$ & $0.78 \pm 0.016$ & $0.755 \pm 0.018$ \\
\hline Re$(y)$ & $0.488 \pm 0.018$ & $0.473 \pm 0.033$ \\
\hline Im$(y)$ & $0.268 \pm 0.004$ & $-0.391 \pm 0.0057$ \\
\hline
\end{tabular}
\]

This model illustrates the predictive potential of models with non-abelian family groups. The $SU(2)$ family symmetry strongly constrains the forms of the mass matrices. 
As noted above, however, the model in its simplest form is not fully realistic. It incorporates the unrealistic ``minimal $SU(5)$" relations between the down quark and charged lepton masses, and it requires some tuning of parameters to explain the mass hierarchy between the first and second families of the down quarks and charged leptons. Nevertheless, the model can account for many qualitative features of the quark and lepton spectrum, and there are ways of repairing some of its defects. For example, the minimal  $SU(5)$ relations can be avoided in several ways, such as introducing Yukawa terms with a ${\bf 45}_H$ or effective dim-5 Yukawa terms with a ${\bf 24}_H$.

\section{Conclusions} 

By decomposition of multiplets of the supergroups $SU(M|N)$, anomaly-free sets of fermion multiplets of the bosonic groups $SU(M) \times SU(N) \times U(1)$ can be found, as was shown in \cite{CFS}. Models based on such groups and multiplets can give both grand unification of the Standard Model gauge interactions and gauged non-abelian family groups. 

In this paper we have explored one potential of such models, namely that the family symmetry could constrain the form of the quark and lepton mass matrices in such a way as to explain the main qualitative features of the quark and lepton properties. We studied the smallest such model that contains three families, which has the group $SU(5) \times SU(2) \times U(1)$. In particular, we studied the minimal form of this model, and showed it can account in a simple way for many of the qualitative features of the spectrum of quark and lepton masses and mixing angles, and make definite predictions for the lepton sector, specifically the Dirac CP phase of the neutrinos. This minimal form is not fully realistic, however, as it gives the ``minimal $SU(5)$" relation between the charged lepton and down quark masses. Nevertheless, the model does illustrate how powerfully the non-abelian family symmetries constrain such models. Very definite and non-trivial forms are obtained for the fermion mass matrices (including that of the right-handed neutrinos), determined in large part by the Clebsch coefficients of the family group. 
 
In addition to their implications for quark and lepton masses and mixing angles, such models in general would have a rich phenomenology if a subgroup of the family gauge groups were broken near the electroweak scale.  This phenomenology would include (a) extra $Z'$ bosons, whose couplings to the quarks and leptons would be quite distinctive; (b) flavor-changing non-abelian gauge interactions, which would give rare flavor-violating decays of leptons, whose branching ratios would be constrained by family symmetry; and (c) extra vector-like quarks and leptons (though these do not appear in the model we studied, due to its small gauge groups). Clearly, there are many possibilities that remain to be explored. Moreover, in such models, one would expect the flavor structure to be sufficiently constrained by the family symmetry to give predictions for proton-decay branching ratios.

\section*{Acknowledgements} The authors acknowledge a discussion with A. Schwimmer. They were partially supported by DOE grant DE-SC0013880.

\end{document}